\documentclass[journal,twoside]{IEEEtran}
\usepackage[latin9]{inputenc}
\usepackage{amsthm}
\usepackage{amstext}
\usepackage{amssymb}
\usepackage{graphicx}
\usepackage{esint}

\makeatletter

\providecommand{\tabularnewline}{\\}

\theoremstyle{plain}
\newtheorem{thm}{\protect\theoremname}

\ifCLASSINFOpdf
\else
\fi
\makeatother

\providecommand{\theoremname}{Theorem}

\begin{document}
\title{Compute and Forward: End to End Performance over Residue Class Based Signal Constellation}

\author{Smrati~Gupta,~\IEEEmembership{Student Member,~IEEE}~and~M.~A.~Vázquez-Castro,~\IEEEmembership{Senior~Member,~IEEE}%
\thanks{Smrati Gupta and M. A. Vázquez-Castro are with the Department of Telecommunications
and Systems Engineering, Universitat Autonoma de Barcelona, Barcelona,
08193, Spain e-mail: smrati.gupta@uab.es, angeles.vazquez@uab.es.%
}%
\thanks{Manuscript received Decmber xx, 2012; revised January xx, 2013.%
}}

\maketitle
\markboth{IEEE COMMUNICATION LETTERS,~Vol.~x, No.~x, January~20xx}{Gupta \MakeLowercase{\textit{et al.}}:Compute and Forward: End to End Performance over Residue Class Based Signal Constellation}

\begin{abstract}
In this letter, the problem of implementing compute and forward (CF)
is addressed. We present a practical signal model to implement CF
which is built on the basis of Gaussian integer lattice partitions.
We provide practical decoding functions at both relay and destination
nodes thereby providing a framework for complete analysis of CF. Our
main result is the analytical derivation and simulations based validation
of union bound of probability of error for end to end performance
of CF. We show that the performance is not limited by the linear combination
decoding at the relay but by the full rank requirement of the coefficient
matrix at the destination. \end{abstract}

\section{Introduction}

In wireless networks with multiple users, relaying is an important
technique adopted to maximize the network throughput. In \cite{BNZ},
Nazer and Gastpar proposed a novel strategy of generalized relaying
called Compute and Forward (CF) which enables the relays in any Gaussian
wireless network to decode linear equations of the transmitted symbols
with finite field coefficients, using the noisy linear combinations
provided by the channel. The linear equations in finite field are
transmitted to the destination and upon receiving sufficient linear
equations, the destination can decode desired symbols. Further, information
theoretical tools are used in \cite{BNZ} to obtain the achievable
rate regions. An algebraic approach to implement CF has been introduced
in \cite{Cfeng} where the authors propose to implement CF making
a connection between CF and isomorphism in module theory.

The main contribution of this correspondence is to demonstrate the
implementation of CF using practical signal constellations and study
its end to end performance from source to destination. We use signal
constellations based on one dimensional Gaussian integer lattices
to implement CF. We utilize the natural isomorphism existing between
these signal constellations and finite fields (\cite{Khuber,hardy})
and apply it to design practical encoding and decoding functions at
each node of the system from source to destination. In order to understand
the factors affecting the CF behavior, we consider integral channels.
Therefore, we bypass the errors introduced due to non-integral nature
of the channel thereby avoiding the ``self-noise'' \cite{BNZ}.
We show that at high SNR, the overall performance of CF is determined
primarily by the choice of the finite field used and is not limited
by the detection of linear combinations at the relay. We also provide
a tight union bound estimate of probability of error at the destination
of CF.

\section{Preliminaries : Gaussian Integers}

In this section, we will present some useful algebraic preliminaries
relevant to this letter. Details can be found in \cite{Khuber,hardy}. 

Let $\mathcal{G}$ be the Gaussian Integers $\mathbb{Z}[i]$ and let
$\mathcal{G}_{\pi}$ denote the residue class $\mathcal{G}$ modulo
$\pi$ where $\pi\in\mathcal{G}$. Any element of $\mathcal{G}$ can
be mapped to the residue class $\mathcal{G}_{\pi}$ using the function
$\mu:\mathcal{G}\rightarrow\mathcal{G}_{\pi}$. which is defined as
\begin{equation}
\mu(g)=g-\left[\frac{g.\pi^{*}}{\pi.\pi^{*}}\right].\pi\label{eq:mu}
\end{equation}
where $\pi^{*}$ is the conjugate of $\pi$, and $\left[.\right]$
is the rounding operation which is defined on complex numbers as $\left[a+bi\right]=\left[a\right]+\left[b\right]i$.
The analogy of $\mathcal{G}$ and $\mathcal{G}_{\pi}$ in integer
domain is $\mathbb{Z}$ and $\mathbb{Z}_{p}$ for some modulo residue
class $\mathbb{Z}\textrm{ mod }p$.

The Gaussian primes are the primes in Gaussian integers which are
given by (i) $\pm1$ and $\pm i$, (ii) the rational primes $p$ with
$p\equiv3\textrm{ mod }4$ and (iii) the factors $a+ib$ of rational
primes $p$ with $p\equiv1\textrm{ mod }4$. The Gaussian primes of
type (iii) exist for every $p\equiv1\textrm{ mod }4$ because the
rational primes of type $p\equiv1\textrm{ mod }4$ can be written
as sum of squares $a^{2}+b^{2}$ by the well known Fermat's Theorem
{[}5, Pg. 291{]}. Therefore,
\[
p=a^{2}+b^{2}=(a+bi)(a-bi)
\]
In this letter, we focus on Gaussian primes of type (iii), although
extension of this work to other types is straight forward.

\section{System Model}

Consider the CF system model with $L$ sources, a relay and a destination
as shown in figure 1. Let $w_{l}\in\mathbb{F}_{p}$ be the message
to be transmitted by the $l$-th source ($l=1,2\ldots L$) chosen
from a finite field $\mathbb{F}_{p}$ of order $p$. The vector of
all the source messages is given by $\mathbf{w}=[\begin{array}{ccc}
w_{1} & \ldots & w_{L}\end{array}]$. Each source encodes the message $w_{l}$ into a complex signal constellation
point using the encoder $\mathcal{E}:\mathbb{F}_{p}\rightarrow\mathbb{C}$
to obtain 
\begin{equation}
x_{l}=\mathcal{E}(w_{l})
\end{equation}
The signals are transmitted across the channel to the relay. In this
model, for the primary understanding, we have assumed that the channel
gains are Gaussian integers and hence there is no ``self-noise''
due to approximation of channel by an integer \cite{BNZ}. It is also
assumed that channel undergoes slow fading and hence remains constant
throughout the transmission of each signal. The signal obtained at
the relay is given by
\begin{equation}
y=h_{1}x_{1}+h_{2}x_{2}+\ldots+h_{L}x_{L}+z\label{eq:y}
\end{equation}
where $h_{l}\in\mathcal{G}$ is the channel coefficient between transmitter
$l$ and the relay node, $z$ is i.i.d Gaussian noise given by $z\sim\mathcal{CN}(0,\sigma^{2})$.
The signal to noise ratio (SNR) is defined as 
\begin{equation}
SNR=\frac{E[\parallel x_{l}\parallel^{2}]}{\sigma^{2}}
\end{equation}
The aim of the relay is to compute a linear combination of source
messages in the original message space $v\in\mathbb{F}_{p}$ given
by
\begin{equation}
v=a_{1}w_{1}\oplus a_{2}w_{2}\ldots a_{L}w_{L}\label{eq:v}
\end{equation}
where $a_{l}\in\mathbb{F}_{p}$ are the linear coefficients chosen
on the basis of $h_{l}$ and $\oplus$ indicates summation over finite
field. The estimate of $v$ obtained at the relay using the decoder
$\mathcal{D}_{R}:\mathbb{C}\rightarrow\mathbb{F}_{p}$ is given by$ $
\begin{equation}
\hat{v}=\mathcal{D}_{R}(y)
\end{equation}
The estimate of the linear combination $\hat{v}$ is transmitted to
the destination. Here we assume this transmission between relay to
destination is error free and the linear combination is obtained at
the destination exactly as estimated at the relay. The destination
obtains $L$ such linear combinations. Therefore, the decoder at the
destination is given by $\mathcal{D}_{D}:\left\{ \mathbb{F}_{p}\right\} ^{L}\rightarrow\left\{ \mathbb{F}_{p}\right\} ^{L}$
such that 
\[
\mathbf{\hat{w}}=\mathcal{D}_{D}(\mathbf{\hat{v}})
\]
where $\mathbf{\hat{w}}$ is the estimate of the $L$ original source
signal vector $\mathbf{w}$ and $\mathbf{\hat{v}}$ is the vector
of estimates of the $L$ linear combinations.
\begin{figure}
\includegraphics[width=8cm,height=8cm,keepaspectratio]{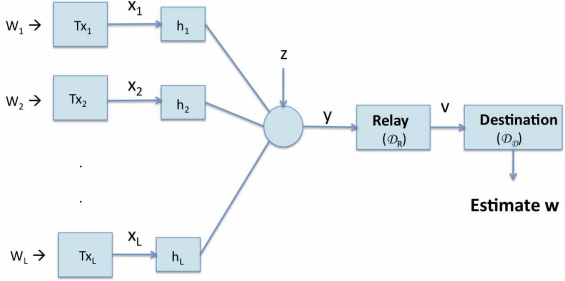}

\caption{End to End CF System Model}
\end{figure}

\section{Proposed Encoding and Decoding Functions}

In this section, we propose the encoding function for the sources
and the decoding functions at the relay and the destination in order
to implement CF scheme.

\subsection{Construction of the Signal Constellation}

We define some standard useful functions \cite{Khuber} which we utilize
in constructing the signal constellations to implement CF. A signal
constellation feasible to implement CF is desired to be isomorphic
to a finite field. Therefore, a natural choice is the residue class
of Gaussian integers $\mathcal{G}_{\pi}$ because any residue class
$\mathcal{G}_{\pi}$ is isomorphic to a finite field $\mathbb{F}_{p}$
if $\pi$ is a prime in $\mathcal{G}$. The size of the field is given
by $p=\mid\pi\mid^{2}$. This isomorphism is defined by the bijective
function $\varphi:\mathbb{F}_{p}\rightarrow\mathcal{G}_{\pi}$ defined
as
\begin{equation}
\varphi(a)=\xi=a-\left[\frac{a.\pi^{*}}{p}\right].\pi\label{eq:psi}
\end{equation}
and the inverse $\varphi^{-1}:\mathcal{G}_{\pi}\rightarrow\mathbb{F}_{p}$
given by
\begin{equation}
a=\varphi^{-1}(\xi)=\xi.(v\pi^{*})+\xi^{*}(u\pi^{*})\textrm{ mod }p\label{eq:psi_inv}
\end{equation}
where $u.\pi+v\pi^{*}=1$ and the Euclidean algorithm can be applied
to calculate $u$ and $v$. With this isomorphism, $\mathcal{G}_{\pi}$
and $\mathbb{F}_{p}$ are mathematically equivalent. In figure 2,
some examples of residue class $\mathcal{G}_{\pi}$ along with their
finite field mapping are shown.  We will now propose the encoding
and decoding functions at the sources, relay and destination.

\begin{figure}
\begin{centering}
\includegraphics[width=8cm,height=5cm,keepaspectratio]{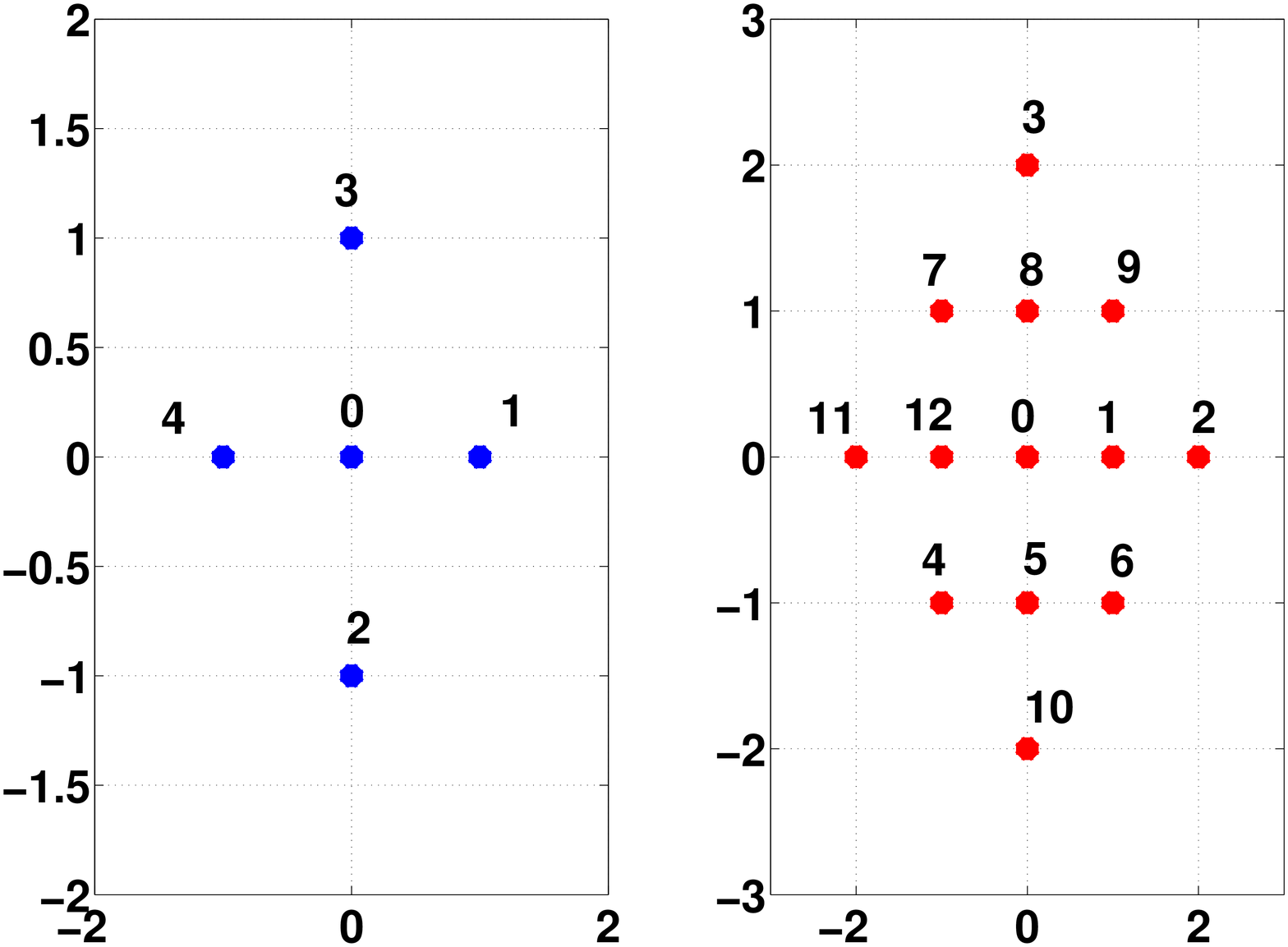}
\par\end{centering}

\caption{Constellations from residue class $\mathcal{G}_{\pi}$ for $\pi=2+i$
and $\pi=3+2i$ and their respective mapping to finite fields $\mathbb{F}_{5}$
and $\mathbb{F}_{13}$}
\end{figure}

\subsection{Encoding at the source}

Let $W$ be the message space which is a finite field comprising of
$p$ elements such that $W=\mathbb{F}_{p}$ . The source messages
are chosen from the message space $w_{l}\in W$. This message space
is required to be isomorphic to some complex signal constellation
$S$ in order to implement CF. The encoding at the source is therefore
done as follows:

1. Choose a signal space size as $\pi=p^{1/2}$ where $\pi\in\mathcal{G}$.
The signal space is hence given by $S=\mathcal{G}_{\pi}$.

2. For each $w_{l}\in W$, obtain the isomorphic element in $\mathcal{G}_{\pi}$
using the bijection function in (\ref{eq:psi}) as $\varphi:W\rightarrow S$
such that 
\[
x_{l}=\varphi(w_{l})
\]
The encoded signals are transmitted to the relay where a noisy linear
combination of the signals is obtained given by (\ref{eq:y}). In
the next subsection, we discuss the decoding performed at the relay.

\subsection{Decoding at the relay}

The relay aims to compute the linear combination\linebreak{}
 $v\in W=\mathbb{F}_{p}$, 
\[
v=a_{1}w_{1}\oplus a_{2}w_{2}\ldots a_{L}w_{L}
\]
where $a_{l}$ is the finite field mapping of the channel gain $h_{l}\in\mathcal{G}$
given by 
\begin{equation}
a_{l}=\varphi(\mu(h_{l}))\label{eq:al}
\end{equation}
Particularly, $h_{l}$ is firstly mapped to the residue class $\mathcal{G}_{\pi}$
using the function $\mu$ defined in (1) and then mapped to finite
field using $\varphi$ in (\ref{eq:psi}). The decoding process at
the relay comprises of the following steps:

1. From the received signal $y$, obtain a maximum likelihood (ML)
estimate of $y$
\begin{equation}
\hat{y}_{ML}=\arg\min_{t\in\mathcal{G}}\parallel y-t\parallel^{2}\label{eq:y_ML}
\end{equation}

2. Map the ML estimator output with the corresponding residue class
element in $S=\mathcal{G}_{\pi}$ using (1) as 
\begin{equation}
\hat{u}=\mu(\hat{y}_{ML})\label{eq:u_ML}
\end{equation}
The output of this operation yields $\hat{u}\in\mathcal{G}_{\pi}$
which is the estimate of linear combination in signal space domain.

3. Map the estimated signal constellation point $\hat{u}$ to message
space given by finite field $W=\mathbb{F}_{p}$ using (\ref{eq:psi_inv})
to obtain 
\begin{equation}
\hat{v}=\varphi^{-1}(\hat{u})\label{eq:v_ML}
\end{equation}
The output of this operation yields an estimate of the linear combination
of the original source signals in finite field $W$.

An error occurs at the relay if the linear combination is incorrectly
estimated. More precisely, the probability of error at the relay is
\begin{equation}
P_{R}=\Pr(\hat{v}\neq v)\label{eq:pe_r}
\end{equation}
The relay transmits the estimate of the linear combination to the
destination where the original source signals are decoded.

\subsection{Decoding at the destination}

The destination collects $L$ linear combinations from the relay which
can be written as
\begin{equation}
\underbrace{\left[\begin{array}{c}
\hat{v}^{1}\\
\vdots\\
\hat{v}^{L}
\end{array}\right]}_{\hat{\mathbf{v}}}=\underbrace{\left[\begin{array}{ccc}
a_{1}^{1} & \ldots & a_{L}^{1}\\
\vdots &  & \vdots\\
a_{1}^{L} & \ldots & a_{L}^{L}
\end{array}\right]}_{\mathbf{A}}\underbrace{\left[\begin{array}{c}
w_{1}\\
\vdots\\
w_{L}
\end{array}\right]}_{\mathbf{w}}\label{eq:matV}
\end{equation}
where $\hat{v}^{t}$ denotes the $t-$th linear combination ($t=1\ldots L$)
and $a_{l}^{t}$ denotes the $l-$th coefficient in $t-$th linear
combination between the $l$th source and relay given by $ $(\ref{eq:al}).
The decoder at the destination inverts the matrix $\mathbf{A}$ and
obtains an estimate of $\mathbf{w}$. Therefore, 
\[
\mathbf{\hat{w}}=\mathbf{A^{-1}}\hat{\mathbf{v}}
\]
Note that here the inverse of $\mathbf{A}$ is taken in $\mathbb{F}_{p}$
and $\mathbf{A}$ is required to be full rank in $\mathbb{F}_{p}$
for successful decoding.

The probability of error at the destination is given by 
\begin{equation}
P_{D}=\Pr(\mathbf{\hat{w}}\neq\mathbf{w})\label{eq:pr_D}
\end{equation}
Therefore, an error occurs at the destination if the original signals
are incorrectly estimated.

\section{Probability of Error}

In this section, we derive an analytical expression for probability
of error at the destination. Since the probability of error at the
destination is also dependent on the probability of error at the relay,
therefore, the later is consequently derived.

Recall from equation (\ref{eq:pr_D}) that the probability of error
at the destination is the probability of decoding incorrect original
source signals such that $P_{D}=\Pr(\hat{\mathbf{w}}\neq\mathbf{w})$.
Therefore, there is an error in detection of $\mathbf{w}$, if there
is an error at the relay in computing any of the $L$ linear combinations
of original signals or if all the \emph{L} linear combinations are
not independent (and consequently, $\mathbf{A}$ in (\ref{eq:matV})
is not full rank). In the next theorem , we present a theoretical
expression for the union bound on the probability of error at the
destination.
\begin{thm}
The union bound estimate of probability of error at the destination
in CF with L sources using finite field of size p and Gaussian integer
residue class based signal constellation is given by 
\[
P_{D}\leq P_{1}+(LP_{R})
\]
where 
\[
P_{1}=1-\prod_{t=1}^{L}\left(1-\frac{1}{p^{t}}\right)
\]
and 
\[
P_{R}=1-\left(erf\left(\frac{1}{2\sqrt{2}\sigma}\right)\right)
\]
such that $\sigma^{2}$ is the variance of additive noise at the relay.\end{thm}
\begin{IEEEproof}
An error occurs at the destination if there is an error in detection
of any linear combination at the relay node and/or the linear combinations
at the destination are not independent (and consequently, $\mathbf{A}$
is not full rank). Therefore, the union bound estimate of probability
of error is given by 
\[
P_{D}\leq P_{1}+\sum_{L}P_{R}
\]
where $P_{1}$ is the probability of $\mathbf{A}$ to have a rank
failure (in $\mathbb{F}_{p}$) and $P_{R}$ is the probability of
error at the relay.  It has been proved in \cite{det} that the probability
of an $L\times L$ matrix $\mathbf{A}$ over a finite field of size
$p$, not being full rank is given by
\begin{equation}
P_{1}=\Pr(\mid\mathbf{A}\mid=0)=1-\prod_{t=1}^{L}\left(1-\frac{1}{p^{t}}\right)\label{eq:p1}
\end{equation}
To evaluate the probability of error at the relay, we use the classic
notion of estimation of error probability. Recall from equation (\ref{eq:pe_r})
that the probability of error at the relay is the probability of decoding
an incorrect linear combination such that $P_{R}=\Pr(\hat{v}\neq v)$.We
rewrite $\hat{v}$ using (\ref{eq:y_ML})-(\ref{eq:v_ML}) as $\hat{v}=\varphi^{-1}(\mu(\hat{y}_{ML})))$.
Since the maps $\mu$ and $\varphi$ are discrete, the equation (\ref{eq:pe_r})
can be written as 
\[
P_{R}=\Pr(\hat{y}_{ML}\neq(h_{1}x_{1}+h_{2}x_{2}+\ldots h_{L}x_{L}))
\]
Since $h_{l}$, $x_{l}$ $\in\mathcal{G}$, therefore, the above expression
is reduced to the probability that the added noise exceeds the voronoi
region of $\mathcal{G}$. The noise is assumed to have a Gaussian
distribution with mean 0 and variance $\sigma^{2}$. Hence,%
\footnote{Since noise has Gaussian distribution, $P_{R}=\Pr\left(\parallel z\parallel>\frac{1}{2}\right)=1-\left(\frac{1}{\sqrt{2\pi\sigma^{2}}}\intop_{-1/2}^{1/2}e^{-\frac{u^{2}}{2\sigma^{2}}}du\right)$,
and the result follows.%
}
\begin{equation}
P_{R}=erfc\left(\frac{1}{2\sqrt{2}\sigma}\right)\label{eq:p_relay}
\end{equation}
where $erfc(x)=\frac{2}{\sqrt{\pi}}\intop_{x}^{\infty}e^{-t^{2}}dt$. 

Further, the probability of error in decoding $L$ linear combinations
at the relay is given by $\sum_{L}P_{R}=LP_{R}$ because all the transmissions
are considered independent. Inserting $P_{1}$ and $P_{R}$ in union
bound estimate, the result is proved.
\end{IEEEproof}
It is clear from (\ref{eq:p1}) that the probability of rank failure
is dependent on the number of users $L$ and the finite field size
$p$ whereas probability of error at the relay (\ref{eq:p_relay})
is dependent only on the additive noise.

\section{Performance Analysis}

In this section, we present the simulations to illustrate the performance
of the proposed encoding and decoding functions in terms of (i) the
probability of error at the relay, which measures error in detecting
linear combinations, (ii) the probability of error at the destination,
which measures the probability of incorrect detection of original
signals. We consider $L=2$ users sending out signals to the destination
via relay. We study the performance of our scheme using different
residue classes $\mathcal{G}_{\pi}$ and their corresponding finite
fields $\mathbb{F}_{p}$. These classes have been listed in Table
I giving the residue class, corresponding fields and the $u$ and
$v$ values to design the isomorphism $\varphi$ in (\ref{eq:psi})-(\ref{eq:psi_inv}).
Further, we consider uniformly distributed channel gains between all
the nodes. For each residue class, we make $L\times10^{4}$ transmissions
from source to destination and the decoding of original signals is
done after every $L$ transmissions. 

Figure 3 shows the comparison of probability of error with varying
SNR. It can be seen that a higher order finite field (or a higher
order $\mathcal{G}_{\pi}$) gives a higher probability of error at
the relay for the same SNR. This happens because the source of error
at the relay is only the additive noise. The impact of this additive
noise is determined by packing and a higher order field will have
a denser packing as compared to lower order field for same SNR.

\begin{figure}
\includegraphics[width=9cm,height=9cm,keepaspectratio]{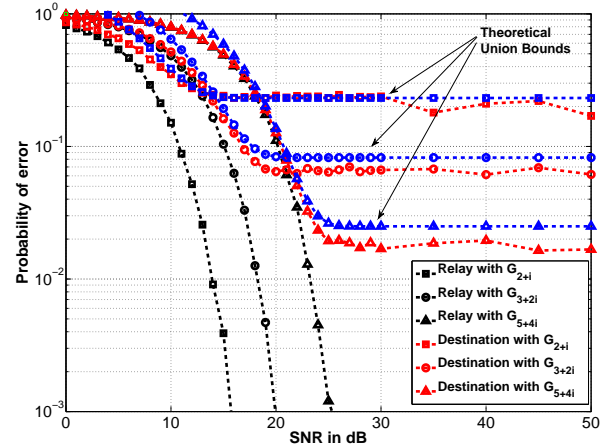}

\caption{Probability of error at the relay and at the destination. In all cases,
$L=2$ users are considered. Three different partitions are plotted.}
\end{figure}
However, at the destination, the probability of error decreases with
increasing SNR up to a certain point and then it attains a constant
value. This is because the overall error is contributed not only by
the additive noise at the relay but also due to the probability of
rank failure at the destination. The probability of rank failure is
independent of SNR (\ref{eq:p1}) and is fixed for any given field
size and number of users. The probability of error at the destination
decreases with increasing SNR only up to the point when it becomes
comparable to the probability of rank failure for a given field size.
After this point, the error at the relay becomes negligible as compared
to error due to rank failure and therefore, error probability at the
destination becomes a constant equal to rank failure probability.
A higher order partition gives a lower probability of error at the
destination at high SNR due to lower probability of rank failure as
compared to lower order partition like $\mathcal{G}_{2+i}$. Also,
note that the theoretical union bound estimate given in Theorem 1
is reasonably tight.

\section{Conclusions}

In this letter, we have introduced a concrete scheme to implement
Compute and Forward relaying protocol using finite size signal constellations.
We have designed encoding and decoding functions using residue class
of Gaussian integers and used their natural properties of isomorphism
with finite fields to obtain mapping between signal space and message
space. We have obtained an analytical union bound estimate of probability
of error and validated it via simulations. We proved that at high
SNR, full rank requirement of the coefficient matrix plays the key
role in determining the end to end performance of CF.
\begin{table}
\begin{centering}
\begin{tabular}{|c|c|c|c|}
\hline 
$p$ & $\pi$ & $u$ & $v$\tabularnewline
\hline 
\hline 
$5$ & $2+i$ & $-1$ & $1+i$\tabularnewline
\hline 
$13$ & $3+2i$ & $-2$ & $1+2i$\tabularnewline
\hline 
$41$ & $5+4i$ & $-4$ & $1+4i$\tabularnewline
\hline 
\end{tabular}
\par\end{centering}

\caption{Finite fields $p$ (where $p\equiv1\textrm{ mod }$ $4$), $\pi$
(where $p=\pi\pi^{*}$) and the values of $u,v$ (where $u\pi+v\pi^{*}=1$)}
\end{table}

\end{document}